\documentclass[aip,jap,reprint,nobalancelastpage]{revtex4-1}
\usepackage{epsf}
\usepackage[T1]{fontenc}
\usepackage[utf8]{inputenc}
\usepackage{amstext}
\usepackage{graphicx}
\usepackage{amssymb}
\usepackage{multirow}
\usepackage{commath}
\usepackage{amsmath,amssymb}
\usepackage{verbatim}


\begin{document}
\def \Bcapp {\text{B}^{\textrm{cool}}_{\textrm{applied}}}
\def \Qtlsparticipation {1/Q_{\text{TLS}} = \Sigma_{i}\, p_i \tan \delta_i}

\preprint{}

\title{Dielectric surface loss in superconducting resonators with flux-trapping holes}

\author{B. Chiaro}
\affiliation{Department of Physics, University of California, Santa Barbara, California 93106, USA}
\author{A. Megrant}
\affiliation{Department of Physics, University of California, Santa Barbara, California 93106, USA}
\affiliation{Department of Materials, University of California, Santa Barbara, California 93106, USA}
\author{A. Dunsworth}
\affiliation{Department of Physics, University of California, Santa Barbara, California 93106, USA}
\author{Z. Chen}
\affiliation{Department of Physics, University of California, Santa Barbara, California 93106, USA}
\author{R. Barends}
\affiliation{Google Inc., Santa Barbara, California 93117, USA}
\author{B. Campbell}
\affiliation{Department of Physics, University of California, Santa Barbara, California 93106, USA}
\author{Y. Chen}
\affiliation{Google Inc., Santa Barbara, California 93117, USA}
\author{A. Fowler}
\affiliation{Google Inc., Santa Barbara, California 93117, USA}
\author{I.C. Hoi}
\affiliation{Department of Physics, University of California, Santa Barbara, California 93106, USA}
\author{E. Jeffrey}
\affiliation{Google Inc., Santa Barbara, California 93117, USA}
\author{J. Kelly}
\affiliation{Google Inc., Santa Barbara, California 93117, USA}
\author{J. Mutus}
\affiliation{Google Inc., Santa Barbara, California 93117, USA}
\author{C. Neill}
\affiliation{Department of Physics, University of California, Santa Barbara, California 93106, USA}
\author{P. J. J. O'Malley}
\affiliation{Department of Physics, University of California, Santa Barbara, California 93106, USA}
\author{C. Quintana}
\affiliation{Department of Physics, University of California, Santa Barbara, California 93106, USA}
\author{P. Roushan}
\affiliation{Google Inc., Santa Barbara, California 93117, USA}
\author{D. Sank}
\affiliation{Google Inc., Santa Barbara, California 93117, USA}
\author{A. Vainsencher}
\affiliation{Department of Physics, University of California, Santa Barbara, California 93106, USA}
\author{J. Wenner}
\affiliation{Department of Physics, University of California, Santa Barbara, California 93106, USA}
\author{T. C. White}
\affiliation{Department of Physics, University of California, Santa Barbara, California 93106, USA}
\author{John M. Martinis}

\affiliation{Department of Physics, University of California, Santa Barbara, California 93106, USA}
\affiliation{Google Inc., Santa Barbara, California 93117, USA}

\date{\today}

\begin{abstract}

Surface distributions of two level system (TLS) defects and magnetic vortices are limiting dissipation sources in superconducting quantum circuits.  Arrays of flux-trapping holes are commonly used to eliminate loss due to magnetic vortices, but may increase dielectric TLS loss.  We find that dielectric TLS loss increases by approximately 25\,\% for resonators with a hole array beginning 2\,$\mu \text{m}$ from the resonator edge, while the dielectric loss added by holes further away was below measurement sensitivity.  Other forms of loss were not affected by the holes.  Additionally, we estimate the loss due to residual magnetic effects to be \mbox{$9\times 10^{-10} /\mu\text{T} $} for resonators patterned with flux-traps and operated in magnetic fields up to $5$\,$\mu\text{T}$.  This is orders of magnitude below the total loss of the best superconducting coplanar waveguide resonators.

\end{abstract}
\maketitle

Superconducting coplanar waveguide (SCPW) resonators are extensively used in astronomy\cite{day2003,mazin2012} and quantum information \cite{mariantoni2011,barends2013,jeffrey2014}.  An important frontier in SCPW resonator development is increasing the intrinsic quality factor $Q_i = 1/\text{loss}$.  This is an especially important proxy for qubit performance, since the resonator $Q_i$ is strongly correlated with the qubit relaxation time $T_1$ because qubits and resonators are subject to many of the same dissipation mechanisms.\cite{wang2015, wisbey2010, song2009a, wang2014, nsanzineza2014, martinis2005, martinis2009, gao2008}  Quantum computers require small operating temperatures $\lesssim 100\,\textrm{mK}$, single-photon excitation energies, low magnetic fields $\lesssim 5\mu\textrm{T}$, and high coherence  $Q_i \gtrsim 10^{6}$.  In this quantum computing regime, dominant loss mechanisms are two-level state (TLS) defects in amorphous dielectrics located at surfaces and loss from trapped flux in magnetic vortices.  

In this Letter, we examine the tradeoff between increased TLS loss and reduced magnetic vortex loss that occurs when the ground plane of SCPW resonators is patterned with an array of holes.  Fractal resonators also reduce magnetic losses, but are typically optimized for use in high magnetic fields and have not demonstrated quality factors as high as coplanar designs in small field environments.\cite{graaf2012}  Although hole arrays have long been known to eliminate dissipation from trapped flux, \cite{song2009b, bothner2011, bothner2012} these structures have not been studied in the quantum computing regime for the possibility of increasing TLS loss. Our data shows that dielectric TLS loss from flux-trapping holes is an important physical limitation if designed incorrectly.   

When a thin-film superconductor is cooled through its transition temperature $T_c$ in a magnetic field $B_{\textrm{cool}}$, it is energetically favorable for magnetic flux to be trapped as vortices at some defect\cite{stan2004}.  The typical spacing between vortices or an edge of the superconducting film to a vortex is $\left( \Phi_{0} / B_{\textrm{cool}} \right)^{1/2}$. As the superconducting order parameter has to vanish\cite{tinkham}, this normal core produces dissipation in response to currents flowing past the core\cite{song2009a}.  With a hole in the film, vortices form without a normal core and produce no dissipation.  We note that suitably positioned normal-core vortices may be beneficial as quasiparticle traps\cite{nsanzineza2014,wang2014}.  For this application, hole arrays should be positioned properly to engineer the number and position of the normal-core vortices. 

Because the holes have sharp edges and expose the substrate, they introduce new dissipation sites from surface TLS defects.  As modern high-$Q$ resonators are sensitive to  nanometer thick amorphous dielectrics at surfaces\cite{wenner2011}, these additional edges can increase loss if the holes are placed near the resonator where the electric fields are the largest.  Consequently, we must determine how closely holes can be safely placed from the resonator.

\begin{figure}
\begin{center}
\includegraphics[width=250 pt]{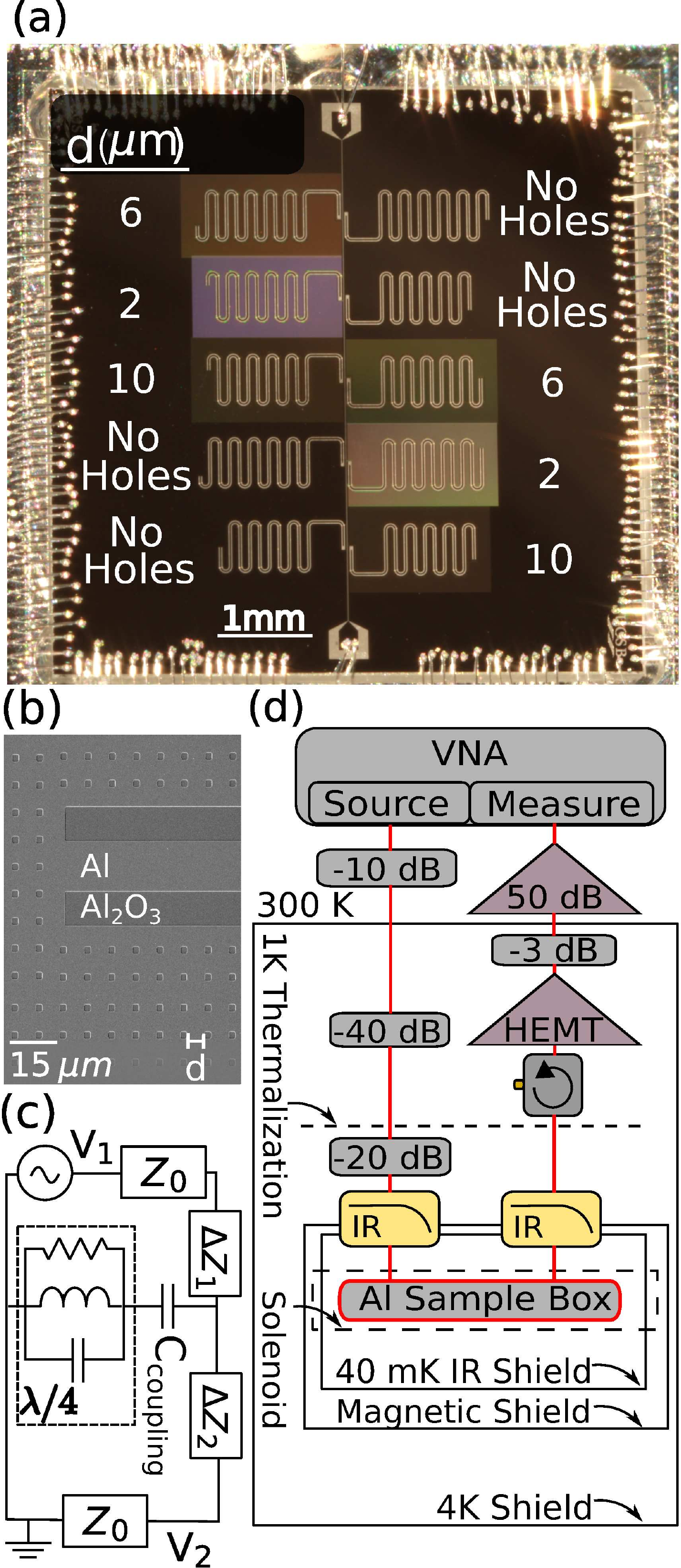}

\end{center}
\caption{(color) Device and apparatus.  (a) Optical micrograph of a chip wirebonded inside sample mount, showing hole edge to edge separation d.  (b) Scanning electron microscope (SEM) image showing a $\lambda / 4$ resonator with a ground-plane hole array, displayed near the antinode of current.  (c) The equivalent circuit for a $\lambda$/4 resonator capacitively near resonance\cite{megrant2012}, coupled to a transmission line.  Included is the effect of small in-line impedance asymmetries characterized by $\Delta Z_{1}$ and $\Delta Z_{2}$.  (d)  Apparatus diagram and wiring schematic with signal path in red.}

\label{Schematic}
\end{figure}

We characterize this loss with quarter-wavelength SCPW resonators that are capacitively coupled to a feedline, with frequency multiplexing to measure 10 resonators per chip.  An optical image of a device wirebonded in a mount is shown in Fig.\,\ref{Schematic}(a). The resonators have fundamental frequencies between 4.6\,GHz and 5.5\,GHz and center trace and gap dimensions of $15\,\mu\text{m}$ and $10\,\mu\text{m}$.  Our circuit contains both resonators with and without ground-plane holes for direct comparison.  Our arrays are made from square holes of side length $2\,\mu\text{m}$ and an edge to edge separation $d$ of $2\,\mu \text{m}$, $6\,\mu \text{m}$, or $10\,\mu \text{m}$.  The distance $d$ is also the distance between the edge of the resonator gap and the nearest hole.\cite{chiaro2015supp}  An example is shown in Fig.\,\ref{Schematic}(b).  The equivalent circuit diagram for this device near resonance is shown in Fig.\,\ref{Schematic}(c) and was analysed in detail in Ref.\,\onlinecite{megrant2012}.

Our resonator circuits were fabricated from 100\,nm aluminium thin films grown on c-plane sapphire substrates. The first type is made in a conventional electron beam deposition system with base pressure of $~3\times 10^{-8}\,\textrm{Torr}$.  Films from this tool yield resonators with $Q_i \simeq 8 \times 10^{5}$ near a measurement photon number $N_\textrm{photon} = 1$ and are thus representative of resonators made with standard deposition techniques.  The second type was prepared in a molecular beam epitaxy (MBE) system with a base pressure of $~ 1\times 10^{-11}\,\textrm{Torr}$.  With an in-situ $\text{O}_2$ plasma cleaning of the substrate at $650$\,$^\circ$C \cite{megrant2012}, we found lower resonator loss $Q_i \simeq 1.5 \times 10^{6}$ for $N_\textrm{photon}  = 1$.  The high quality factors make the MBE grown resonators sensitive probes of subtle decoherence mechanisms that may be induced by the holes. We report measurements from two circuits from each film for a total of four circuits.  The resonators and holes were etched simultaneously in an inductively coupled plasma.  The etch was performed at 0.7 Pa using BCl$_3$ and Cl$_2$ flow rates of 20 and 40 SCCM and 70 W bias power.

For measurement, individual devices were wirebonded in Al sample mounts and anchored to the cold stage of an adiabatic demagnetization refrigerator (ADR).  A schematic of our apparatus is shown in Fig.\,\ref{Schematic}(d).  The ~40\,mK ADR base temperature is well below the transition temperature $T_c \simeq 1.1\,\textrm{K}$ of Al, so the thermal quasiparticle density is negligible.  Additionally, our cryostat includes extensive infra-red (IR) radiation shielding composed of in-line coaxial IR filters and a light tight sample compartment that reduces the non-equilibrium quasiparticle population below our measurement sensitivity \cite{barends2011}.  A circulator on the output line of the chip reduces noise from the input of the high electron mobility transistor (HEMT) amplifier. A solenoid encircles the sample compartment allowing us to apply a magnetic field perpendicular to the film, with $50\,\textrm{nT}$ resolution to measure the magnetic field dependence of our resonator $Q_i$.  We surround the mount with a magnetic shield and remove all magnetic components.  We test each component that we use inside the sample compartment for magnetism and use non-magnetic SMA connectors (EZ Form Cable Corp. model \#705626-301), cables (EZ Form Cable Corp. model \#301844), brass screws, and custom experimental hardware.  This reduces the ambient magnetic field at the device to $\lesssim 1.5\,\mu\textrm{T}$.

\begin{figure*}
\begin{center}

\includegraphics[width=500 pt]{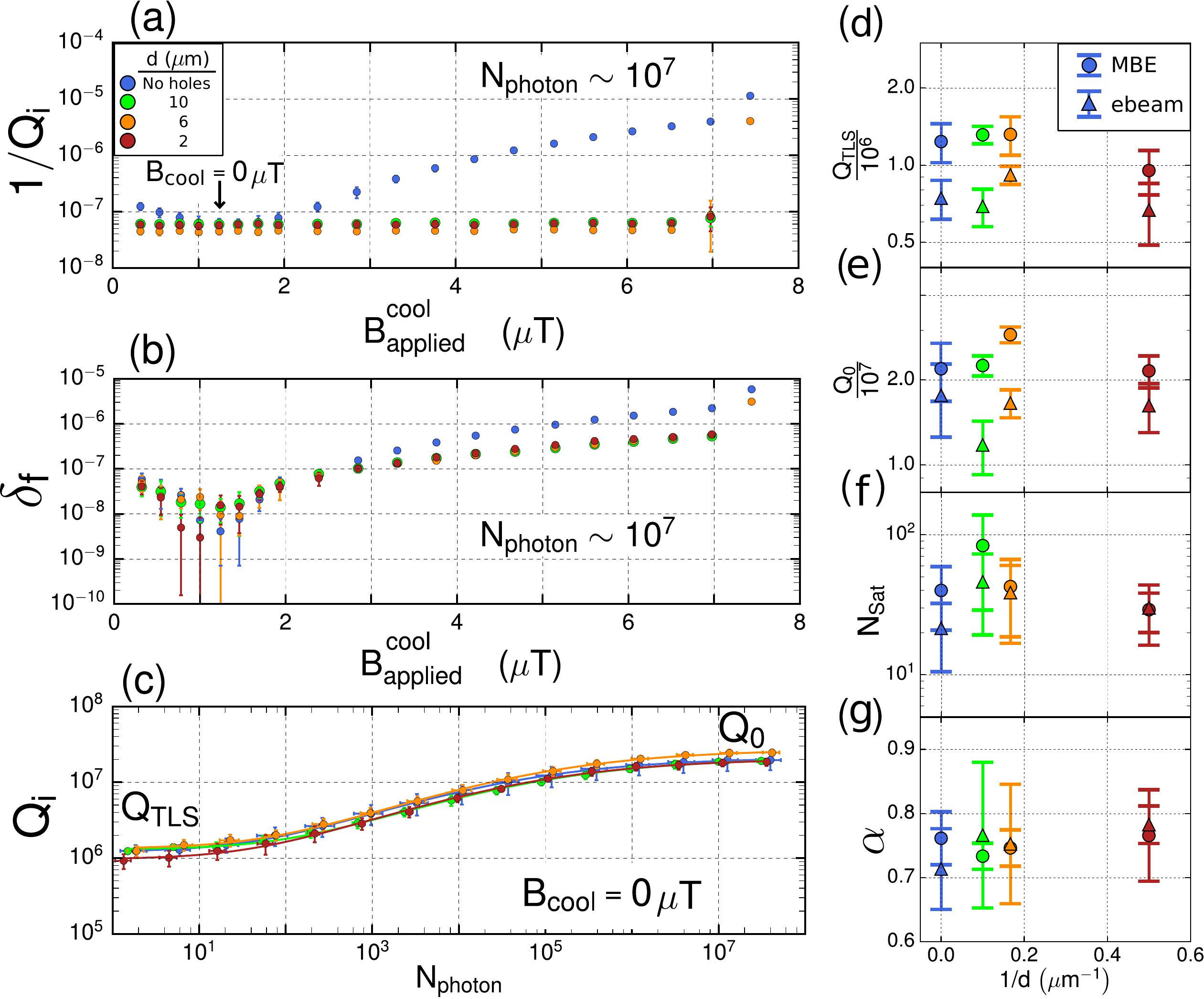}

\end{center}
\caption{(Color) Measurement results  (a) and (b) Loss $1/Q_i$ and resonator fractional frequency shift $\delta_{\textrm{f}}$ versus applied magnetic field when cooled through $T_c$ $\Bcapp$, taken at high power $N_\textrm{photon} \sim 10^7$.  Data is for the MBE-grown film and hole patterns of varying density.  The resonators with no holes have the greatest sensitivity to magnetic fields, so the minimum of loss identifies the true zero giving the cryostat offset field.  (c) The dependence of $Q_i$ on measurement drive power for the MBE sample, fit to a standard TLS dissipation model from Eqn.\,\ref{TLS_dissipation}.  (d) - (g) Extracted TLS model parameters from power dependence measurements as shown in (c) for the MBE and e-beam samples, showing loss attributed to unsaturated TLS (low power) and due to power independent mechanisms (high power) vs the edge to edge hole spacing d.  In all plots the data shows mean value for resonators of a common hole density, with error bars indicating 1 standard deviation.}
\label{Resonator Measurements}
\end{figure*}

The loss is determined by measuring the resonator intrinsic quality factor $Q_i$ using transmission spectroscopy\cite{megrant2012}.  In the initial measurement phase, we measure $Q_i$ versus applied magnetic field at high power for better signal to noise ratio, since vortex loss has weak power dependence\cite{bothner2011}.  To vary $B_\textrm{cool}$, we raise the device temperature above $\text{T}_{c}$, set the applied magnetic field $\Bcapp$, and cool the sample back through its $\text{T}_{c}$ in this field thereby trapping magnetic vortices.  Once the device has returned to its base temperature we extract the resonator $Q_i$ from $S_{21}$ measurements with a vector network analyzer (VNA).

Figure\,\ref{Resonator Measurements} (a) shows the magnetic field dependence of $Q_i$ for resonators with and without ground plane holes from the MBE device.  For resonators without a patterned ground plane there is a well defined maximum of $Q_i$ that identifies the applied field that zeros the total magnetic field.\cite{chiaro2015supp}  The offset between $\Bcapp$ and $B_{\textrm{cool}}$ is indicated by the arrow.  We observe a gradual but significant increase in loss away from this field that is attributed to a greater density of magnetic vortices trapped in the ground plane.  As expected, for resonators with holes we find that $Q_i$ is nearly independent of applied magnetic field until the critical field for vortex formation in the center trace has been exceeded.  Figure\,\ref{Resonator Measurements} (b) shows the fractional frequency shift  \mbox{$\delta_{\text{f}}=( f_{0}(\Bcapp=0) - f_0(\Bcapp))/f_{0}(\Bcapp=0)$} we observe that resonators with patterned ground planes are less susceptible to field induced frequency shifts.

For MBE grown resonators with hole patterns we consider data in the field range of $\Bcapp=1.4-6.4$\,$\mu\text{T}$.  Fields in this range are less than the critical field for vortex formation in the center trace of the resonator and allows us to estimate the residual magnetic loss in the absence of local magnetic vortices.  By assuming an excess loss model that is linear in $\Bcapp$, we estimate the residual magnetic loss to be \mbox{$8.6 \pm 1.3 \times 10^{-10}$\,/$\mu\text{T}$.}  We show the data supporting this estimate in the supplement.\cite{chiaro2015supp}  Although additional experiments are required to determine the origin and proper functional dependence of this excess loss, we suggest likely models are coupling to vortices in remote areas of the device not protected by the hole overlays or quasiparticles generated by the local suppression of $T_c$ due to the magnetic field.  For typical shielded devices, this estimate is several orders of magnitude below the loss of the best SCPW resonators.\cite{megrant2012, ohya2014, bruno2015}

After measuring the magnetic field dependence of the high power $Q_i$ we quantify the surface loss from TLS defects by measuring the power dependence of the resonator $Q_i$.  We use the value of $\Bcapp$ that maximizes $Q_i$ at high power, so that $B_{\text{cool}} = 0\,\mu\text{T}$ as described previously.  The power dependence data for the MBE device is shown in Fig.\,\ref{Resonator Measurements}(c), where the lines are fits to a standard TLS loss model \cite{wang2009}
\begin{equation}
\label{TLS_dissipation}
\frac{1}{Q_i} = \frac{1}{Q_\text{TLS}}\frac{1}{\sqrt{1+\left( \frac{N_\text{photon}} {N_\text{sat}}\right)^\alpha}} + \frac{1}{Q_0}
\end{equation}
This model decomposes the total internal loss of the resonator $1/Q_i$ into a power independent loss term $1/Q_0$ that includes such loss modes from quasiparticles and radiation, and a power dependent term of magnitude $1/Q_\text{TLS}$ that comes from TLS defects.  Here $N_\text{photon}$ is the excitation number of photons in the resonator and $N_\text{sat}$ describes the saturation field of the TLS bath.  The parameter $\alpha$ is related to the electric field distribution of the resonator, and may be influenced by interactions between TLS defects within the bath.\cite{faoro2012, faoro2015}

Figure \ref{Resonator Measurements}(d) and (e) show the quality factors for the low power ($Q_\textrm{TLS}$) and high power ($Q_0$) regimes, extracted from the fits in (c).  The data points represent resonators from two circuits from each film.  In (d), we see that the densest hole pattern increases TLS loss by roughly 25\% relative to resonators without holes for both the MBE and ebeam grown resonators.  Dielectric TLS loss is often decomposed as \mbox{$\Qtlsparticipation$} where \mbox{$\tan \delta_i$} is the loss tangent of dielectric volume $i$ and the participation ratio $p_i$ is the fraction of the electric energy of the resonator excitation that is stored within that volume.\cite{wang2015, barends2010, gao2008}  We presume that the increase in TLS loss is not due to an increase in the dielectric loss tangent, but rather due to an increase in the participation ratio resulting from a redistribution of the electric field.

To quantify the excess loss due to the dense hole pattern we perform linear regression analysis controlling for the difference between the MBE and ebeam films.\cite{chiaro2015supp}  This analysis includes the results from 23 resonators, 11 measured on two circuits from the MBE film and 12 measured on two circuits from the ebeam film.  We find that the TLS loss $1/Q_\text{TLS}$ directly attributable to the dense hole pattern is $2.5 \pm 1.3\times 10^{-7}$, where the uncertainty represents the standard error.  The p-value for this regression coefficient is 0.07.

When using the resonators for quantum devices at low magnetic fields, this increase in loss is undesirable.  The hole spacing should thus be carefully chosen, first to be close enough to provide protection from external fields of magnitude $\sim \Phi_0/d^2$, where $d$ is the edge to edge spacing between holes\cite{stan2004}.  However, the spacing from the resonator to the first row of holes should be greater than about $6\,\mu\textrm{m}$, a value that did not exhibit measurable excess TLS loss.  In (e) we observe that the $Q_0$ of resonators with the densest hole pattern is nearly the same as that without any hole overlay.  This indicates that power independent loss mechanisms were not affected by the hole patterns.

We have characterized dissipation from arrays of flux-trapping holes in SCPW resonators.  We find that excess dielectric loss can be made vanishingly small by increasing the distance between the resonator edge and the array.  In our experiment, a 6\,$\mu \text{m}$ separation was enough to remove excess dielectric loss; power-independent loss mechanisms were not affected by the arrays.  We also estimate the residual magnetic loss in resonators with ground plane holes to be $\sim$\,$10^{-9}$\,/$\mu\text{T}$, showing that SCPW resonators can be made insensitive to small magnetic fields without magnifying other loss mechanisms.

\section*{ACKNOWLEDGEMENTS}%
Devices were fabricated at the UCSB Nanofabrication Facility, a part of the NSF-funded National Nanotechnology Infrastructure Network.  This research was funded by the Office of the Director of National Intelligence (ODNI), Intelligence Advanced Research Projects Activity (IARPA), through Army Research Office Grant No. W911NF-09-1-0375 and by Google Inc. All statements of fact, opinion, or conclusions contained herein are those of the authors and should not be construed as representing the official views or policies of IARPA, the ODNI, the U.S. Government.

\clearpage
\newcommand{\beginsupplement}{%
        \setcounter{table}{0}
        \renewcommand{\thetable}{S\arabic{table}}%
        \setcounter{figure}{0}
        \renewcommand{\thefigure}{S\arabic{figure}}%
        \setcounter{equation}{0}
        \renewcommand{\theequation}{S\arabic{equation}}%
     }
\beginsupplement
\makeatletter
\makeatother

\begin{center}
\textbf{Supplementary information for "Dielectric surface loss in superconducting resonators with flux-trapping holes"}
\end{center}

\def \Bcapp {\text{B}^{\text{cool}}_{\text{applied}}}

We show the data and analysis estimating residual magnetic loss in superconducting resonators with flux-trapping hole arrays. We describe the design rules used to embed a meandered resonator in a rectangular array of holes.  We show data supporting the claim that the maximum in $Q_{i}$ is obtained at an applied magnetic field that cancels the ambient field.  We report the TLS model parameters for resonators from the MBE and ebeam deposited films from which we estimate the excess loss attributable directly to the densest hole pattern.
\maketitle

In the main text, we infer resonator loss 1/$Q_i$ by measuring the system scattering parameters (S-parameters) of superconducting coplanar waveguide (SCPW) resonators capacitively coupled to a feedline.  The circuit model for this system has been analysed previously \cite{megrant2012} and gives the result 
\begin{equation}
\label{spectroscopyS21}
\tilde{S}_{21}^{-1} = 1 + \frac{Q_{i}}{Q_{c}^{*}}e^{i \phi}\frac{1}{1+i2 Q_{i} \frac{f-f_{0}}{f_{0}}}
\end{equation}
Here $\tilde{S}_{21}^{-1}$ is the inverse of transmission data calibrated to enforce $S_{21}=1$ far off resonance.  $f_{0}$ and $\phi$ are the resonant frequency and impedance mismatch angle and  $Q_{c}^{*}$ is the coupling quality factor scaled by an impedance ratio.

\begin{figure}[b]
\begin{center}
\includegraphics[width=240 pt]{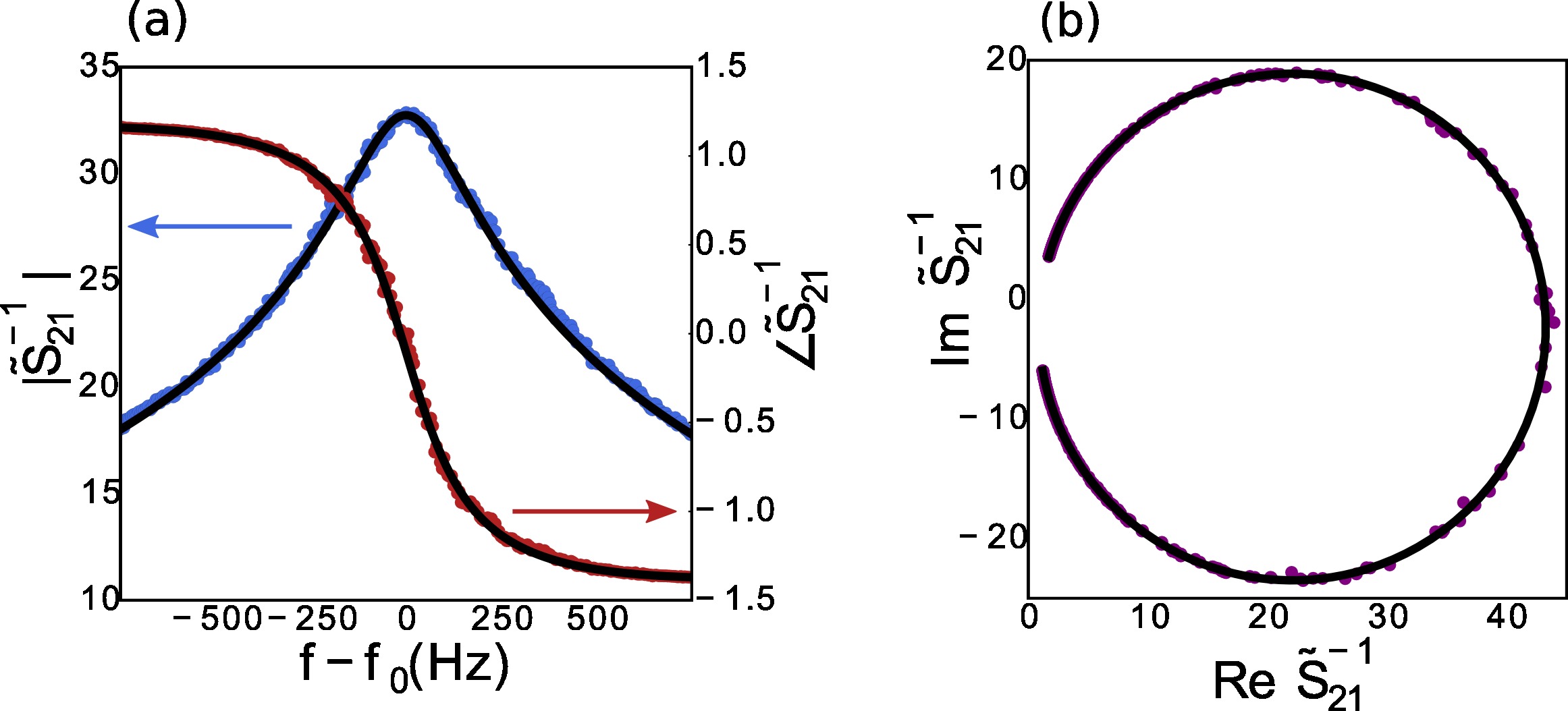}
\caption{(Color online)  The transmission spectroscopy data for the extraction of $Q_{i}$ shown as (a) magnitude and phase and (b) in the complex plane.}
\label{basicfit}
\end{center}
\end{figure}

\begin{figure}
\begin{center}
\includegraphics[width=230 pt]{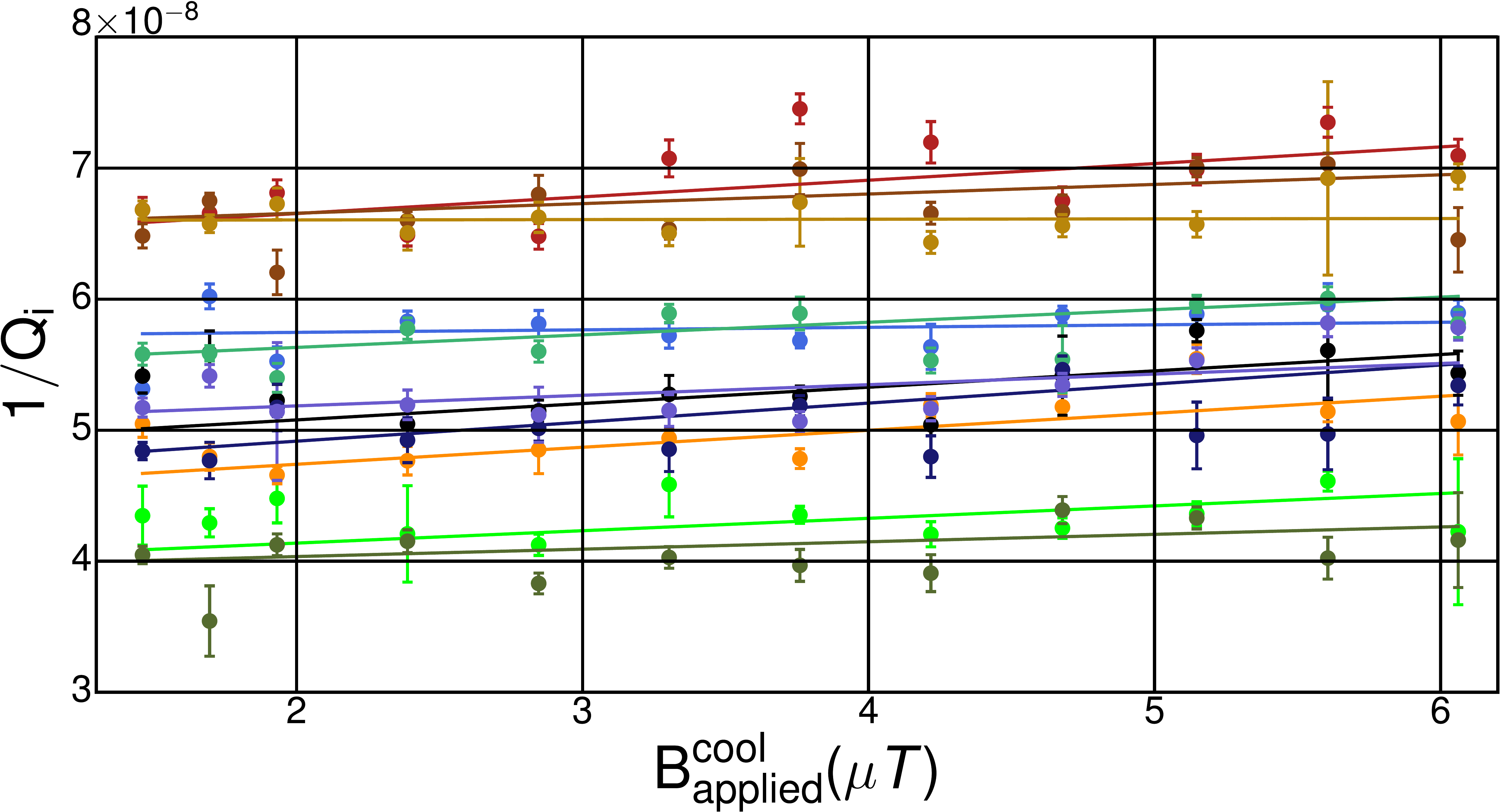}
\caption{(Color)  The small magnetic field dependence of internal loss for 11 resonators with ground plane hole arrays.}
\label{novortexdata}
\end{center}
\end{figure}

We extract the internal quality factor $Q_{i}$ and its statistical uncertainty $\sigma_{Q_{i}}$ by fitting measurements of our device to this model.  Figure \ref{spectroscopyS21} shows example data from the MBE sample for which the $Q_{i} = 1.86 \times 10^{7} \pm 2.58 \times 10^{5}$.  This is representative of the data points shown in Fig\,\ref{novortexdata}.

Fig.\,\ref{novortexdata} breaks out the data in Fig.\,2a of the main text for resonators with ground plane holes from the MBE sample in the region from $\Bcapp$ = 1.4\,$\mu T$ to 6.4\,$\mu T$.  This is below the critical field for vortex formation in both the center trace of the resonator and in the ground plane near the resonator where there are ground plane holes.  Thus, we do not expect magnetic loss in this region.  In Fig.\,\ref{novortexdata} we apply a linear loss model $1/Q_{i}$\,=\,$m$\,$\times$\,$\Bcapp$\,+\,$b$ to the data and obtain the parameters m and b with a weighted least squares fit.  We use weighting factors $1/\sigma^2_{\left( 1/Q_i \right)}$, where $\sigma_{\left( 1/Q_i \right)}=\left( -1/Q_{i}^{2} \right) \sigma_{Q_{i}}$ in terms of parameters extracted directly from device data.  We obtain our estimate of residual magnetic loss from the parameter $m$ in these fits.  This estimate is $m = 8.6\times10^{-10} \pm 1.3\times 10^{-10}$ /$\mu T$, where the uncertainty represents the standard error.

\setlength{\textfloatsep}{0pt}

\begin{figure}
\begin{center}
\includegraphics[width=240 pt]{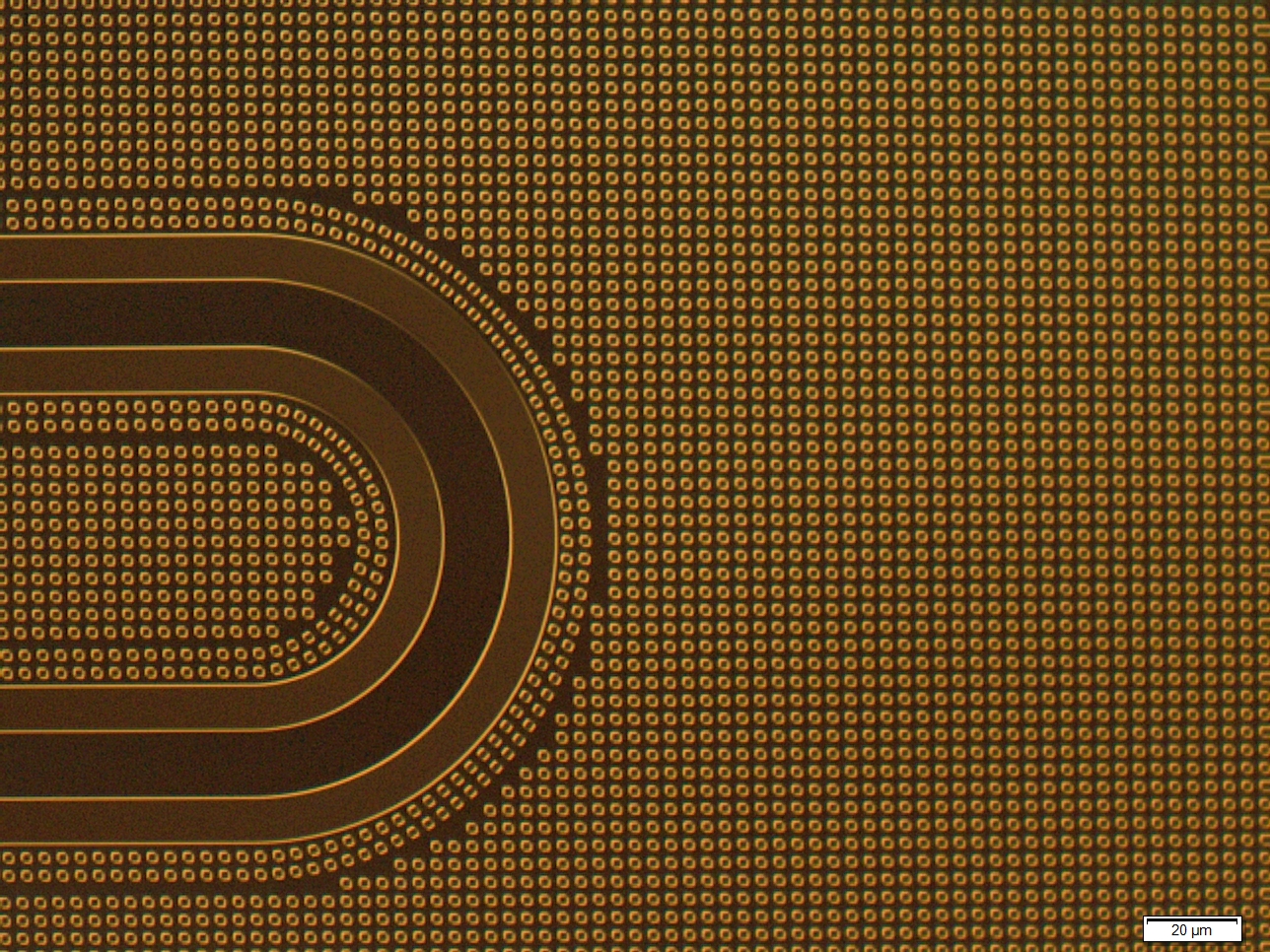}
\caption{The meandered section of a $\lambda$/4 resonator with a flux trapping holes showing two rows holes following the contour of the meander and the rectangular array of holes in which the resonator is embedded.}
\label{meander_holearray}
\end{center}
\end{figure}

In the main text, we describe the hole pattern as having a constant distance between the edges of adjacent holes and between the resonator edge and the nearest hole.  However, the curved portion of our meandered resonator geometry does not allow us to easily satisfy both of these constraints simultaneously.  We expect that excess dielectric loss from hole patterns will be primarily determined by the resonator edge to hole edge distance and dominated by the row of holes nearest the resonator edge.  The compromise that we have adopted is to have two rows of holes follow the contour of the resonator with a constant spacing between adjacent holes and the resonator edge.  This composite structure is then embedded in a rectangular array of regularly spaced holes.  An example is shown in fig. \ref{meander_holearray}.  This allows us to maintain a constant resonator to hole edge distance for the entire length of the resonator.

\begin{figure}
\begin{center}
\includegraphics[width=240 pt]{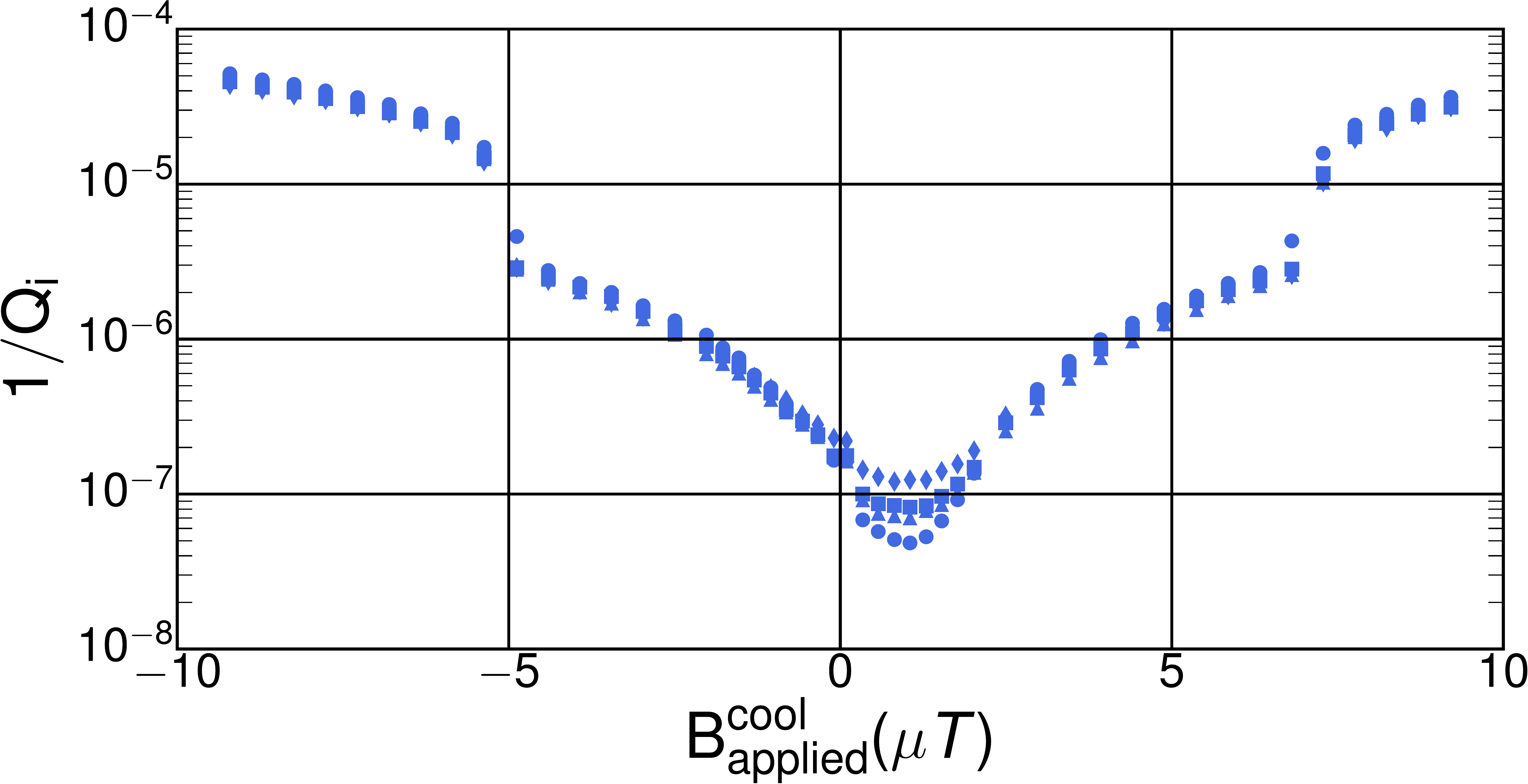}
\caption{The full field dependence of $Q_i$ for four resonators without flux-trapping holes from the ebeam deposited film.  The single loss minimum is observed when the total magnetic field is zero.}
\label{fullfield}
\end{center}
\end{figure}

In the main text, we claim that the maximum in $Q_i$ is obtained at an applied magnetic field that cancels the ambient field.  If the $Q_i$ maximum were due to vortex mediated quasiparticle recombination\cite{nsanzineza2014} rather than field cancellation then we would expect to observe a second maximum at a symmetric, negative magnetic field.   Fig\,\ref{fullfield} shows the field dependence of $Q_i$ at both positive and negative fields for four resonators without flux-trapping holes on one circuit from the ebeam deposited film.  The absence of a second maximum at negative fields shows that vortex mediated quasiparticle recombination does not play an important role in determining $Q_{i}$ in our system.  This is expected because our cryostat features extensive shielding and filtering to reduce the number of nonequilibrium quasiparticles and we operate the resonators well below $T_c$ so that the number of thermal quasiparticles is also small. 

\setlength{\textfloatsep}{0pt}

In the main text, we claim that the TLS loss due to the densest hole pattern is $2.5 \pm 1.3\times 10^{-7}$.  This claim is derived from measurements of 23 resonators summarized in table \ref{table:ResonatorData}.  We use the R statistics package\cite{R} to apply a linear regression to this data set, accounting for both the deposition technique and the hole pattern.  No interaction term was included in this model because the interaction between hole pattern and deposition technique was found to be statistically insignificant.  This analysis yields an estimate for TLS loss due to the dense hole pattern, the standard error of that estimated value, and a p-value indicating the statistical significance of this finding.  Those values were respectively $2.5\times10^{-7}$, $1.3\times10^{-7}$, and 0.07.

\begin{table}
\caption{Deposition condition, hole density, and TLS model parameters extracted from the power dependence at $\Bcapp$ = 0}
\centering
\begin{tabular}{ c c c c c c}
\hline \hline
Deposition & 1/d $(\mu m^{-1})$ & $1/\text{Q}_{\text{tls}}$ & $1/\text{Q}_{0}$ & $\text{N}_{\text{sat}}$ & $\alpha$ \\
\hline
MBE & 0 & 7.3e-07 & 3.55e-08 & 21.8 & 0.81 \\ 
MBE & 0 & 6.59e-07 & 4.92e-08 & 35.4 & 0.73 \\ 
MBE & 0 & 1.19e-06 & 5e-08 & 10.7 & 0.71 \\ 
MBE & 0 & 7.44e-07 & 5.82e-08 & 71.3 & 0.73 \\ 
MBE & 0 & 7.3e-07 & 3.32e-08 & 54.1 & 0.82 \\ 
MBE & 0 & 8.73e-07 & 4.44e-08 & 34.1 & 0.79 \\ 
MBE & 0 & 9.4e-07 & 6.64e-08 & 52.7 & 0.75 \\ 
MBE & 0.5 & 8.9e-07 & 4.01e-08 & 24.6 & 0.71 \\ 
MBE & 0.5 & 9.26e-07 & 5.29e-08 & 16.4 & 0.68 \\ 
MBE & 0.5 & 1.03e-06 & 4.23e-08 & 39.1 & 0.81 \\ 
MBE & 0.5 & 1.56e-06 & 5.35e-08 & 36.4 & 0.86 \\ 
ebeam & 0 & 1.1e-06 & 3.77e-08 & 20.1 & 0.75 \\ 
ebeam & 0 & 1.26e-06 & 5.61e-08 & 9.7 & 0.69 \\ 
ebeam & 0 & 1.2e-06 & 7.21e-08 & 34.8 & 0.74 \\ 
ebeam & 0 & 1.63e-06 & 8.55e-08 & 37.3 & 0.80 \\ 
ebeam & 0 & 1.88e-06 & 4.18e-08 & 8.0 & 0.78 \\ 
ebeam & 0 & 1.23e-06 & 5.24e-08 & 10.6 & 0.64 \\ 
ebeam & 0 & 1.63e-06 & 5.92e-08 & 20.5 & 0.70 \\ 
ebeam & 0 & 1.18e-06 & 8.75e-08 & 30.5 & 0.61 \\ 
ebeam & 0.5 & 1.06e-06 & 5.25e-08 & 50.8 & 0.82 \\ 
ebeam & 0.5 & 1.54e-06 & 7.52e-08 & 30.8 & 0.80 \\ 
ebeam & 0.5 & 1.55e-06 & 5.06e-08 & 25.8 & 0.74 \\ 
ebeam & 0.5 & 2.3e-06 & 7.83e-08 & 12.6 & 0.77 \\ 
\hline
\end{tabular}
\label{table:ResonatorData}
\end{table}


\begin{thebibliography}{99}

\bibitem{day2003}  Peter K. Day,  Henry G. LeDuc,  Benjamin A. Mazin,  Anastasios Vayonakis, and  Jonas Zmuidzinas, \textit{Nature} \textbf{425},  (2003)
\bibitem{mazin2012}  Benjamin A. Mazin,  Bruce Bumble,  Seth R. Meeker,  Kieran O’Brien,  Sean McHugh, and  Eric Langman, \textit{Optics express} \textbf{20},  (2012)
\bibitem{mariantoni2011}  Matteo Mariantoni, H.  Wang,  Radoslaw C. Bialczak, M.  Lenander,  Erik Lucero, M.  Neeley, A. D.  O’Connell, D.  Sank, M.  Weides, J.  Wenner, T.  Yamamoto, Y.  Yin, J.  Zhao,  John M. Martinis, and A. N.  Cleland, \textit{Nat. Phys.} \textbf{7},  (2011)
\bibitem{barends2013} R.  Barends, J.  Kelly, A.  Megrant, D.  Sank, E.  Jeffrey, Y.  Chen, Y.  Yin, B.  Chiaro, J.  Mutus, C.  Neill, P.  O'Malley, P.  Roushan, J.  Wenner, T. C.  White, A. N.  Cleland, and  John M. Martinis, \textit{Phys. Rev. Lett.} \textbf{111},  (2013)
\bibitem{jeffrey2014}  Evan Jeffrey,  Daniel Sank, J. Y.  Mutus, T. C.  White, J.  Kelly, R.  Barends, Y.  Chen, Z.  Chen, B.  Chiaro, A.  Dunsworth, A.  Megrant, P. J. J.  O'Malley, C.  Neill, P.  Roushan, A.  Vainsencher, J.  Wenner, A. N.  Cleland, and  John M. Martinis, \textit{Phys. Rev. Lett.} \textbf{112},  (2014)
\bibitem{wang2015}Wang, C. and Axline, C. and Gao, Y. Y. and Brecht, T. and Chu, Y. and Frunzio, L. and Devoret, M. H. and Schoelkopf, R. J. \textit{Appl. Phys. Lett.} \textbf{107},  (2015)
\bibitem{wisbey2010}Wisbey, David S. and Gao, Jiansong and Vissers, Michael R. and da Silva, Fabio C. S. and Kline, Jeffrey S. and Vale, Leila and Pappas, David P. \textit{JAP}, \textbf{108},  (2010)
\bibitem{song2009a} C.  Song, T. W.  Heitmann, M. P.  DeFeo, K.  Yu, R.  McDermott, M.  Neeley,  John M. Martinis, and B. L. T.  Plourde, \textit{Physical Review B} \textbf{79},  (2009)
\bibitem{wang2014}  Chen Wang,  Yvonne Y. Gao,  Ioan M. Pop,  Uri Vool,  Chris Axline,  Teresa Brecht,  Reinier W. Heeres,  Luigi Frunzio,  Michel H. Devoret,  Gianluigi Catelani, L. I.  Glazman, and R. J.  Schoelkopf, \textit{Nat. Commun.} \textbf{5},  (2014)
\bibitem{nsanzineza2014} I.  Nsanzineza, and B. L. T.  Plourde, \textit{Phys. Rev. Lett} \textbf{113},  (2014)

\bibitem{martinis2005}John M. Martinis, K. B. Cooper, R. McDermott, Matthias Steffen, Markus Ansmann, K. D. Osborn, K. Cicak, Seongshik Oh, D. P. Pappas, R. W. Simmonds, and Clare C. Yu, \textit{Phys. Rev. Lett} \textbf{95}, (2005)
\bibitem{martinis2009} John M. Martinis, M. Ansmann, and J. Aumentado, \textit{Phys. Rev. Lett} \textbf{103}, (2009)
\bibitem{gao2008} Gao, Jiansong and Daal, Miguel and Vayonakis, Anastasios and Kumar, Shwetank and Zmuidzinas, Jonas and Sadoulet, Bernard and Mazin, Benjamin A. and Day, Peter K. and Leduc, Henry G., \textit{Appl. Phys. Lett.}, \textbf{92}, (2008)

\bibitem{graaf2012} Graaf, S. E. de and Danilov, A. V. and Adamyan, A. and Bauch, T. and Kubatkin, S. E., \textit{JAP}, \textbf{112},  (2012)
\bibitem{song2009b} C.  Song, M. P.  DeFeo, K.  Yu, and B. L. T.  Plourde, \textit{Appl. Phys. Lett.} \textbf{95},  (2009)
\bibitem{bothner2011}  Daniel Bothner,  Tobias Gaber,  Matthias Kemmler,  Dieter Koelle, and  Reinhold Kleiner, \textit{Appl. Phys. Lett.} \textbf{98},  (2011)
\bibitem{bothner2012} D.  Bothner, C.  Clauss, E.  Koroknay, M.  Kemmler, T.  Gaber, M.  Jetter, M.  Scheffler, P.  Michler, M.  Dressel, D.  Koelle, and R.  Kleiner, \textit{Appl. Phys. Lett.} \textbf{100},  (2012)
\bibitem{stan2004}  Gheorghe Stan,  Stuart B. Field, and  John M. Martinis, \textit{Phys. Rev. Lett} \textbf{92},  (2004)
\bibitem{tinkham}Michael Tinkham, Introduction to superconductivity (2012)
\bibitem{wenner2011} J.  Wenner, R.  Barends, R. C.  Bialczak,  Yu Chen, J.  Kelly,  Erik Lucero,  Matteo Mariantoni, A.  Megrant, P. J. J.  O’Malley, D.  Sank, A.  Vainsencher, H.  Wang, T. C.  White, Y.  Yin, J.  Zhao, A. N.  Cleland, and  John M. Martinis, \textit{Appl. Phys. Lett.} \textbf{99},  (2011)
\bibitem{chiaro2015supp}Supplementary information for "Dielectric surface loss in superconducting resonators with flux-trapping holes"
\bibitem{megrant2012} A.  Megrant, C.  Neill, R.  Barends, B.  Chiaro,  Yu Chen, L.  Feigl, J.  Kelly,  Erik Lucero,  Matteo Mariantoni, P. J. J.  O’Malley, D.  Sank, A.  Vainsencher, J.  Wenner, T. C.  White, Y.  Yin, J.  Zhao, C. J.  Palmstrøm,  John M. Martinis, and A. N.  Cleland, \textit{Appl. Phys. Lett.} \textbf{100},  (2012)
\bibitem{barends2011} R.  Barends, J.  Wenner, M.  Lenander, Y.  Chen, R. C.  Bialczak, J.  Kelly, E.  Lucero, P.  O’Malley, M.  Mariantoni, D.  Sank, H.  Wang, T. C.  White, Y.  Yin, J.  Zhao, A. N.  Cleland,  John M. Martinis, and J. J. A.  Baselmans, \textit{Appl. Phys. Lett.} \textbf{99},  (2011)
\bibitem{ohya2014} S. Ohya, B. Chiaro, A. Megrant, C. Neill, R. Barends, Y. Chen, J. Kelly, D. Low, J. Mutus,  P J J O’Malley,  P. Roushan,  D. Sank, A. Vainsencher, J. Wenner, T. C. White, Y. Yin, B. D. Schultz, C. J. Palmstrøm, B. A. Mazin, A. N. Cleland and John M. Martinis,  \textit{Supercond. Sci. Technol.}, (2014)
\bibitem{bruno2015} Bruno, A. and de Lange, G. and Asaad, S. and van der Enden, K. L. and Langford, N. K. and DiCarlo, L., \textit{Appl. Phys. Lett.}, \textbf{106}, (2015)
\bibitem{wang2009} H.  Wang, M.  Hofheinz, J.  Wenner, M.  Ansmann, R. C.  Bialczak, M.  Lenander,  Erik Lucero, M.  Neeley, A. D.  O’Connell, D.  Sank, M.  Weides, A. N.  Cleland, and  John M. Martinis, \textit{Appl. Phys. Lett.} \textbf{95},  (2009)
\bibitem{faoro2012}L. Faoro and L. Ioffe, \textit{Phys. Rev. Lett.}, \textbf{109}, (2012).
\bibitem{faoro2015}L. Faoro and L. Ioffe, \textit{Phys. Rev.B}, \textbf{91}, (2015).
\bibitem{barends2010}  Barends, R. and Vercruyssen, N. and Endo, A. and de Visser, P. J. and Zijlstra, T. and Klapwijk, T. M. and Diener, P. and Yates, S. J. C. and Baselmans, J. J. A., \textit{Appl. Phys. Lett.}, \textbf{97}, (2010)

\end{thebibliography}

\begin{thebibliography}{99}
\bibitem{megrant2012} A.  Megrant, C.  Neill, R.  Barends, B.  Chiaro,  Yu Chen, L.  Feigl, J.  Kelly,  Erik Lucero,  Matteo Mariantoni, P. J. J.  O’Malley, D.  Sank, A.  Vainsencher, J.  Wenner, T. C.  White, Y.  Yin, J.  Zhao, C. J.  Palmstrøm,  John M. Martinis, and A. N.  Cleland, \textit{Appl. Phys. Lett.} \textbf{100},  (2012)
\bibitem{nsanzineza2014} I.  Nsanzineza, and B. L. T.  Plourde, \textit{Phys. Rev. Lett} \textbf{113},  (2014)
\bibitem{R}  R Core Team (2013). R: A language and environment for statistical computing. R Foundation for Statistical Computing, Vienna, Austria.  URL http://www.R-project.org/.

\end{thebibliography}
\end{document}